\documentclass[runningheads]{llncs}

\usepackage{graphicx}
\usepackage{amsmath}
\usepackage{amssymb}
\usepackage{booktabs}

\usepackage{algorithmic}
\usepackage{textcomp}
\usepackage{xcolor}
\usepackage{flushend}
\usepackage{tabularx}
\usepackage{multirow}
\usepackage{array}
\usepackage{booktabs,enumitem}
\usepackage[linesnumbered,ruled,lined,commentsnumbered]{algorithm2e}
\usepackage{color,soul}
\usepackage{tabularx}
\usepackage{enumitem}
\usepackage{eccv}

\usepackage{eccvabbrv}

\usepackage{graphicx}
\usepackage{booktabs}

\usepackage[accsupp]{axessibility}  % Improves PDF readability for those with disabilities.

\usepackage{hyperref}

\usepackage{orcidlink}

\begin{document}
% ---------------------------------------------------------------
% TODO REVIEW: Replace with your title
\title{RMT-BVQA: Recurrent Memory Transformer based Blind Video Quality Assessment for Enhanced Video Content}
% TODO REVIEW: If the paper title is too long for the running head, you can set
% an abbreviated paper title here. If not, comment out.
\titlerunning{RMT-BVQA}

% TODO FINAL: Replace with your author list. 
% Include the authors' OCRID for the camera-ready version, if at all possible.
\author{Tianhao Peng\thanks{Equal contribution}\inst{1} \orcidlink{0009-0008-5294-2880} \and
Chen Feng$^\star$\inst{1}\orcidlink{0009-0001-2480-907X} \and
Duolikun Danier\inst{1}\orcidlink{0000-0002-9320-7099} \and
Fan Zhang\inst{1}\orcidlink{0000-0001-6623-9936} \and
Benoit Quentin Arthur Vallade\inst{2} \and
Alex Mackin\inst{2} \and
David Bull\inst{1}\orcidlink{0000-0001-7634-190X}}

% TODO FINAL: Replace with an abbreviated list of authors.
\authorrunning{T.~Peng et al.}
% First names are abbreviated in the running head.
% If there are more than two authors, 'et al.' is used.

% TODO FINAL: Replace with your institution list.
\institute{Visual Information Laboratory, University of Bristol\\
One Cathedral Square, Bristol, BS1 5DD, United Kingdom \and
Amazon Prime Video, \\
1 Principal Place, Worship Street, London, EC2A 2FA, United Kingdom
}

\maketitle

\begin{abstract}
   With recent advances in deep learning, numerous algorithms have been developed to enhance video quality, reduce visual artifacts, and improve perceptual quality. However, little research has been reported on the quality assessment of enhanced content - the evaluation of enhancement methods is often based on quality metrics that were designed for compression applications. In this paper, we propose a novel blind deep video quality assessment (VQA) method specifically for enhanced video content. It employs a new Recurrent Memory Transformer (RMT) based network architecture to obtain video quality representations, which is optimized through a novel content-quality-aware contrastive learning strategy based on a new database containing 13K training patches with enhanced content. The extracted quality representations are then combined through linear regression to generate video-level quality indices. The proposed method, RMT-BVQA \footnote{The code can be downloaded from: \url{https://jasminepp.github.io/RMTBVQA/}.}, has been evaluated on the VDPVE (VQA Dataset for Perceptual Video Enhancement) database through a five-fold cross validation. The results show its superior correlation performance when compared to ten existing no-reference quality metrics. 
  % \keywords{First keyword \and Second keyword \and Third keyword}
\end{abstract}


\section{Introduction}
\label{sec:intro}

\begin{figure}[!t]
    \centering
    \includegraphics[width=0.95\linewidth]{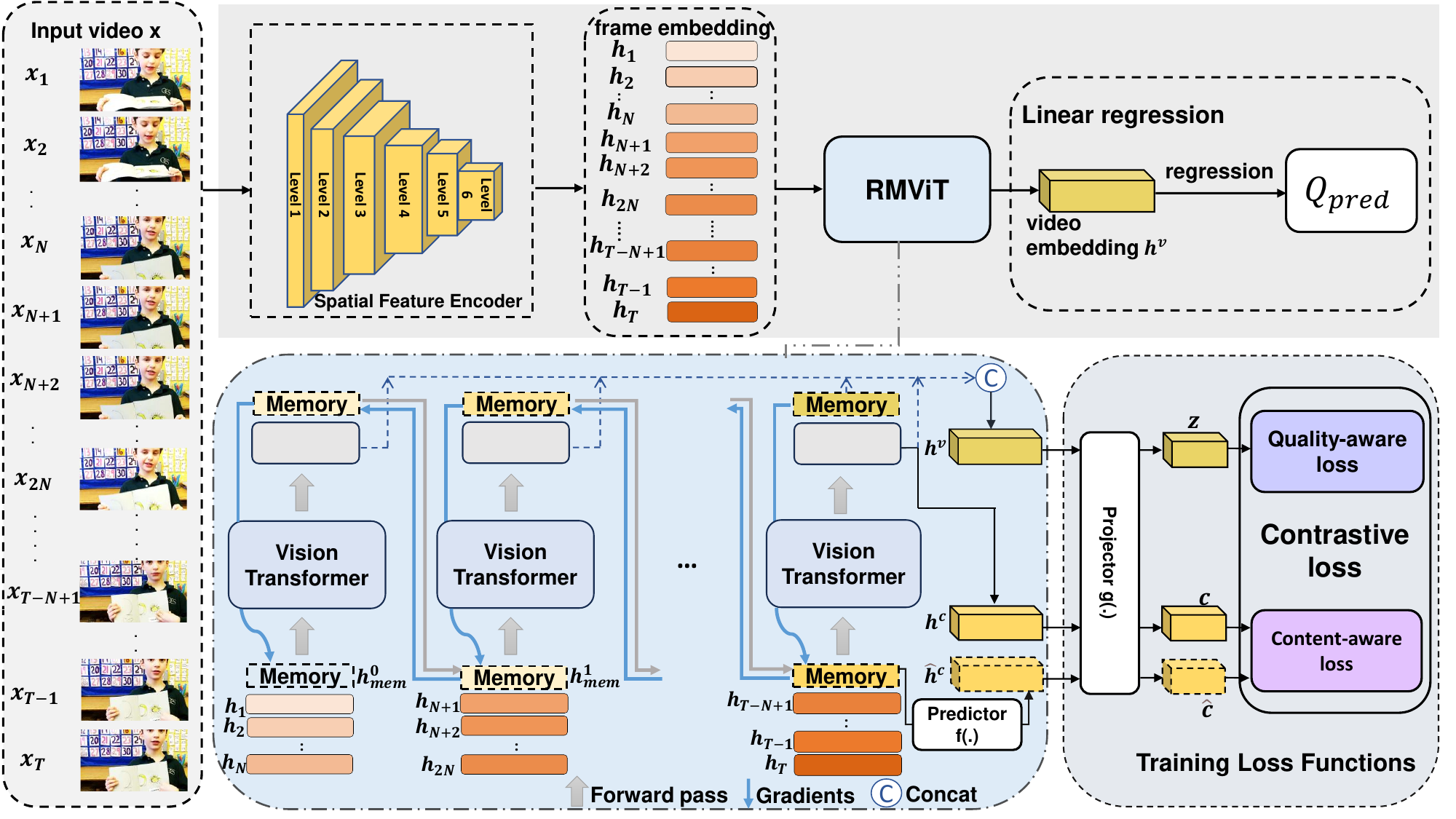}
    \caption{Illustration of the proposed RMT-BVQA framework.}
    \label{fig:framework}
\end{figure}

Video content is now everywhere! It is by far the largest global Internet bandwidth consumer, with a wide range of applications including consumer video, video conferencing, and gaming. It has been reported that in 2022, each individual in the United Kingdom spent 4.5 hours per day (on average) consuming video content on different platforms \cite{ofcom}. Due to various conditions associated with video capture, editing, and delivery, streamed content often contains a range of visual artifacts, which can affect the quality of a user's experience. To address this issue, enhancement approaches have been developed, with the aim of reducing visible artifacts and improving overall perceptual quality, \eg color transform \cite{ACE, MBLLEN}, deblurring \cite{ESTRNN, DeblurGANv2}, deshaking \cite{GlobalFlowNet}, post processing \cite{ViSTRA3, zhang2021video}, super resolution \cite{danielyan2011bm3d, zhang2021vistra2}, etc. In particular, more recently, driven by the advances of deep generative models \cite{ma2024cvegan,danier2024ldmvfi,chen2024hierarchical}, we have seen more effective methods proposed that offer promising enhancement results.

In order to evaluate the performance of these methods, enhanced content can be assessed subjectively through psychophysical experiments or objectively using various video quality metrics. While the former offers ground-truth results, objective video quality assessment (VQA) methods are used more often in practice due to their higher efficiency and lower cost \cite{bull2021intelligent}. In the video enhancement literature, blind (without pristine reference sources) quality metrics are more relevant compared to full- and reduced- reference VQA methods. This is because, in most cases, the enhancement operation is performed at the user end, where the reference content is unavailable. 

Conventional blind VQA methods \cite{brisque, niqe, VIDEVAL} are predominantly based on various features extracted in the spatio-temporal or/and frequency domains. Recent advances in this research area favor deep learning-based models employing convolutional neural networks (CNNs) \cite{DeepVBQA, VSFA} or Vision Transformers (ViTs) \cite{FastVQA}. However, these methods tend to exhibit inconsistent performance due to the lack of large and diverse training databases (in particular, for enhanced video content) and inefficient training methodologies. Furthermore, in order to make use of the limited ground-truth quality labels in existing datasets, end-to-end training at the video level is required, but infeasible due to the computation constraints (e.g. memory) with most existing hardware. As a result, many methods \cite{FastVQA,C3DVQA} resort to subsampling videos before evaluating their quality, leading to a significant loss of information. Moreover, it is also noted that most of these blind VQA models were originally optimized and validated on compressed videos that contain artifact types and (spatio-temporal) distributions that differ compared to enhanced content. This can lead to inconsistent metric performance when used to assess enhanced video content.

Inspired by recent works in contrastive learning \cite{contrique, zhao2023quality} and the recurrent memory mechanism \cite{bulatov2022recurrent}, we proposes a blind VQA method, RMT-BVQA (illustrated in \cref{fig:framework}), based on a new content-quality-aware self-supervised learning methodology and a novel Recurrent Memory Vision Transformer (RMViT) module. This approach first generates a comprehensive video representation through dynamically processing global and local information across video frames using the proposed RMViT module. The video representation is then used to predict the final sequence quality score through linear regression. To facilitate the optimization of the RMViT module through contrastive learning, we have also created a large-scale database containing content generated by various enhancement methods. The primary contributions of this work are summarized as follows. 

\begin{itemize}
     \item[1)] \textbf{Recurrent Memory Vision Transformer (RMViT) module}: To characterize the artifacts exhibiting in the enhanced video content, we designed a new network architecture based on the recurrent memory mechanism \cite{bulatov2022recurrent}, which has been employed in language modeling \cite{bulatov2022recurrent}, for capturing both short-term temporal dynamics and long-term global information. This also aligns well with the visual persistence characteristic of the human visual system \cite{coltheart1980iconic}. Moreover, this facilitates end-to-end training at the video level, without sub-sampling the video content, and the recurrence nature allows our model to evaluate videos of variable length. This is the \textbf{first} time when the Recurrent Memory Vision Transformer\footnote{We acknowledge that other recurrent mechanisms, such as LSTMs, have been employed for VQA in the literature \cite{gotz2021konvid}.} is used for the video quality assessment task.

    \item[2)] \textbf{Content-quality-aware self-supervised learning}: We developed a new self-supervised learning strategy based on a content-quality-aware loss function. This training methodology enables us to optimize the proposed RMViT module based on a large amount of training material without performing expensive and time consuming subjective tests. Here we, for the first time, used a proxy perceptual quality metric to support the quality-aware contrastive learning for VQA, inspired by the ranking-based training strategies  \cite{feng2024rankdvqa,qi2023full}. 
    
     \item[3)] \textbf{A new training dataset}: We developed a large and diverse training dataset containing various types of enhanced video content to support the proposed content-quality-aware contrastive learning strategy. This training dataset is made available for future research.

\end{itemize}

The proposed method has been evaluated on the VQA Dataset for Perceptual Video Enhancement (VDPVE) \cite{gao2023vdpve} and achieves superior performance over ten existing no-reference VQA methods based on five-fold cross validations, with an average Spearman Rank-order Correlation Coefficient (SRCC) of 0.8209.

\section{Related Work}
\label{sec:LR}

\noindent\textbf{Objective quality assessment} is one of the most important research topics in the field of image processing. It aims to accurately predict the perceptual quality of an input signal given (full reference \cite{ssim,w:VMAF,zhang2015perception}) or without (no reference \cite{niqe,Kim_2017_CVPR,FastVQA}) the corresponding reference content\footnote{Another type of quality assessment methods does exist, denoted reduced-reference models, which only employ partial information from the reference.}. Objective quality metrics play an essential role in comparing different image/video processing methods and supporting algorithm optimization, \eg in the rate quality optimization for compression \cite{ndjiki2012perception} or as loss functions for learning-based approaches \cite{ma2024cvegan}. In the context of video enhancement, since the quality assessment of enhanced content is typically performed when the original reference video is absent, we primarily focus on the no-reference scenario here.

\noindent\textbf{Conventional no-reference quality assessment} methods often employ hand-crafted models to observe the input content through spatial, temporal and/or frequency feature extraction. For example, V-BLIINDS  \cite{VBLIINDS} employs a spatio-temporal natural scene statistics (NSS) model to quantify motion coherency in video scenes; TLVQM  \cite{TLVQM} calculates features at various scales from a selection of representative video frames; VIDEVAL \cite{VIDEVAL} predicts video quality by extracting various spatio-temporal artifact features such as motion, jerkiness, and blurriness. Other notable contributions include NIQE \cite{niqe}, BRISQUE \cite{brisque}, and V-CORNIA \cite{V-CORNIA}. A more comprehensive review of blind image and video quality metrics can be found in \cite{shahid2014no}.

\noindent\textbf{Learning-based no-reference quality assessment} has become more popular recently, inspired by advances in machine learning. Early attempts \cite{Kim_2017_CVPR,VBLIINDS,TLVQM,ChipQA} use various regression models to fit and predict quality indices based on extracted features and conventional quality metrics. More recently, deep neural networks have been utilized for both feature extraction and quality regression to offer improved prediction performance, with important examples including BVQA \cite{BVQA}, SimpleVQA \cite{SimpleVQA}, FAST-VQA \cite{FastVQA}, TB-VQA \cite{wu2023video} and SB-VQA \cite{huang2023sb}.

\noindent\textbf{Training methodology} For deep learning-based quality models, it is key to have a large, diverse, and representative training database. However, creating such databases is costly, since ground-truth quality labels are typically annotated through subjective tests involving human participants. To address this issue, Feng \etal \cite{feng2024rankdvqa} used proxy quality metrics for labeling training material and developed a ranking-inspired training strategy to maintain the reliability of quality annotation. Moreover, self-supervised learning methods have also been employed that convert quality labeling into an auxiliary task \cite{contrique, conviqt, zhao2023quality}.

\noindent\textbf{Quality assessment for enhanced video content} is an underexplored research topic. Previous works typically fine-tune existing blind quality models using enhanced content. A grand challenge \cite{liu2023ntire} was organized in 2023 specifically for enhanced video quality assessment, based on a public training database (VDPVE) \cite{gao2023vdpve}, which contains enhanced video sequences generated through contrast enhancement, deshaking and deblurring. However, it should be noted that only the training set in the VDPVE database is publicly available, while the test dataset used in the grand challenge has not been released. 

\section{Proposed Method}
\label{sec: Method}

The proposed RMT-BVQA framework is illustrated in \cref{fig:framework}. It first takes each frame of the input video and transforms it into a one-dimensional embedding using a Spatial Feature Encoder. The extracted embeddings are then {sequentially} processed by the RMViT (Recurrent Memory Vision Transformer) module to generate the representation of the input video. Finally, a linear ridge regression is employed to output the final predicted sequence quality score. The network architectures employed for each module in this framework, the training database, and the model optimization strategy are described below in detail. 

\subsection{Network architecture}

\noindent\textbf{Spatial Feature Encoder.} Here we employ a pre-trained ResNet-50 based \cite{he2016deep} network which was optimized in a deep image quality model \cite{contrique} for spatial feature extraction. The network parameters are fixed during our training process for the proposed RMT-BVQA. For each frame, a 2048$\times$1 embedding is extracted.

\noindent\textbf{RMViT module.} Considering that artifact types and distributions in enhanced video content are different from those in compressed content (on which most blind VQA methods are optimized and validated), we designed a new Recurrent Memory Vision Transformer (RMViT) module to effectively capture both local temporal dynamic and global information within the input video sequence. This is inspired by the recurrent memory mechanism \cite{bulatov2022recurrent} that has been successfully integrated into language models. This mechanism can process and ``remember" both local and global information and pass the ``memory'' between segments within a long sequence using the recurrence structure \cite{bulatov2022recurrent}, thus is applicable to videos of any length. To our knowledge, this is the first time the recurrent memory mechanism has been applied to the quality assessment task.

As shown in \cref{fig:framework}, in the first recurrent iteration, the proposed RMViT module takes $N$ frame embeddings ($\mathbf{h}_1\  \mathbf{h}_{2} \dots \mathbf{h}_{N}$, corresponding to the first segment in the video) extracted by the Spatial Feature Encoder together with empty memory ($\mathbf{h}_{\mathrm{mem}}^0\in \mathbb{R}^{2048\times M}$) as input, where $N$ is a configurable hyperparameter that defines the length of each video segment and $M$ denotes the number of memory tokens. Specifically, for the first iteration, the input of RMViT is given below:
\begin{equation}
    \mathbf{H}_0 = [\mathbf{h}_{\mathrm{mem}}^0 \circ \mathbf{h}_1 \circ \mathbf{h}_{2} \circ ... \circ \mathbf{h}_{N}],
\end{equation}
in which $\circ$ stands for the concatenation operation. $\mathbf{H}_0$ will then be processed by a Vision Transformer with the output, $\tilde{\mathbf{H}}_0\in \mathbb{R}^{2048\times (M+N)}$:
\begin{equation}
     \tilde{\mathbf{H}}_0 := [\mathbf{h}_{\mathrm{mem}}^1 \circ \tilde{\mathbf{h}}_1 \circ \tilde{\mathbf{h}}_{2} \circ ... \circ \tilde{\mathbf{h}}_{N}]= \text{ViT}(\mathbf{H}_0)  
\end{equation}
For the second recurrent iteration, we further combine the frame embeddings of the next video segment ($\mathbf{h}_{N+1}\  \mathbf{h}_{N+2} \dots \mathbf{h}_{2N}$) with the memory token $\mathbf{h}_{\mathrm{mem}}^1$ in $\tilde{\mathbf{H}}_0$ as the input of the Vision Transform $\mathbf{H}_1$ to obtain $\tilde{\mathbf{H}}_1$. 
\begin{equation}
\mathbf{H}_1 = [\mathbf{h}_{\mathrm{mem}}^1 \circ \mathbf{h}_{N+1} \circ \mathbf{h}_{N+2} \circ ... \circ \mathbf{h}_{2N}], \text{ and }
\end{equation}
\begin{equation}
     \tilde{\mathbf{H}}_1 := [\mathbf{h}_{\mathrm{mem}}^2 \circ \tilde{\mathbf{h}}_{N+1} \circ \tilde{\mathbf{h}}_{N+2} \circ ... \circ \tilde{\mathbf{h}}_{2N}]= \text{ViT}(\mathbf{H}_1).  
\end{equation}
After processing all segments in the input video, the RMViT finally outputs a video-level embedding $\textbf{h}^v \in \mathbb{R}^{2048\times1}$ by performing average pooling on the concatenation of the memory tokens generated in the last recurrent iteration and all the Vision Transformer processed frame embeddings, $[\mathbf{h}_{\mathrm{mem}}^S \circ \tilde{\mathbf{h}}_1 \circ \tilde{\mathbf{h}}_{2} \circ ... \circ \tilde{\mathbf{h}}_{T}]$. It is noted that this is slightly different to the original recurrent memory mechanism \cite{bulatov2022recurrent}, where only the memory in the last iteration contributes to the module output. This modification helps to capture the temporal dynamics across the entire video sequence and enhances the stability of the model. Here $S$ stands for the number of segments, while $T$ represents the total number of frames in $S$ segments. In cases when the last segment contains frames fewer than $N$, this segment will be discarded.

\noindent\textbf{Regression.} The final output of the RMViT module, $\textbf{h}^v$, is passed to a linear ridge regressor (the same as in \cite{conviqt}) in order to obtain the predicted sequence level quality score, $Q_{pred}$. Here, the regressor is optimized during the cross-validation experiment, in which the model parameters of the RMViT are fixed.

\subsection{Training Database Generation}
\label{ssec:DatabaseGeneration}

 \begin{table}[!t]
     \caption{Training database generation.\label{tab:database}}
    \begin{center}
    \small
    \begin{tabular}{m{3.5cm}|m{2cm}|m{6cm}}
    \toprule
    Class & Number of Source Videos &Enhancement Methods\\ \midrule 
    Colour, Brightness and Contrast Enhancement & \multicolumn{1}{c|}{44}   & ACE \cite{ACE}, DCC-Net \cite{DCC-Net}, MBLLEN \cite{MBLLEN}, CapCut \cite{CapCut} \\\midrule 
    Deshaking &  \multicolumn{1}{c|}{50}  & GlobalFlowNet \cite{GlobalFlowNet}, Adobe Premiere Pro \cite{Adobe},  CapCut (minimum cropping mode) \cite{CapCut}, CapCut (most stable mode) \cite{CapCut} \\\midrule 
    Deblurring & \multicolumn{1}{c|}{55}  & ESTRNN \cite{ESTRNN}, DeblurGANv2 \cite{DeblurGANv2}, BasicVSR++ \cite{BasicVSR++}, Adobe Premiere Pro \cite{Adobe}\\
    \bottomrule  
    \end{tabular}
    \end{center}
    \end{table}

To support the training of the RMViT module, we have collected 149 source videos (with a spatial resolution of 1080p or 720p) from five publicly available datasets including BVI-DVC \cite{BVI-DVC}, KoNViD-1k \cite{KoNViD-1k}, LIVE-VQC \cite{livevqc}, Live-Qualcomm \cite{live-qualcomm}, and YouTube-UGC \cite{YouTube}. Three primary visual artefacts\footnote{The investigation of other types of enhanced content was not the focus of this work, but remains as future work.} (as defined in VDPVE \cite{gao2023vdpve}) including (i) color / brightness / contrast degradation, (ii) camera shaking, and (iii) blurring are synthesized (for (iii), through Gaussian filtering) or inherent within the source content (for (i) and (ii)). These videos are then processed using 12 different conventional and learning-based enhancement methods to generate a total of 596 enhanced video sequences. The detailed database generation process is summarized in \cref{tab:database}.

\begin{figure}[t!]
    \centering
    \includegraphics[width=0.95\linewidth]{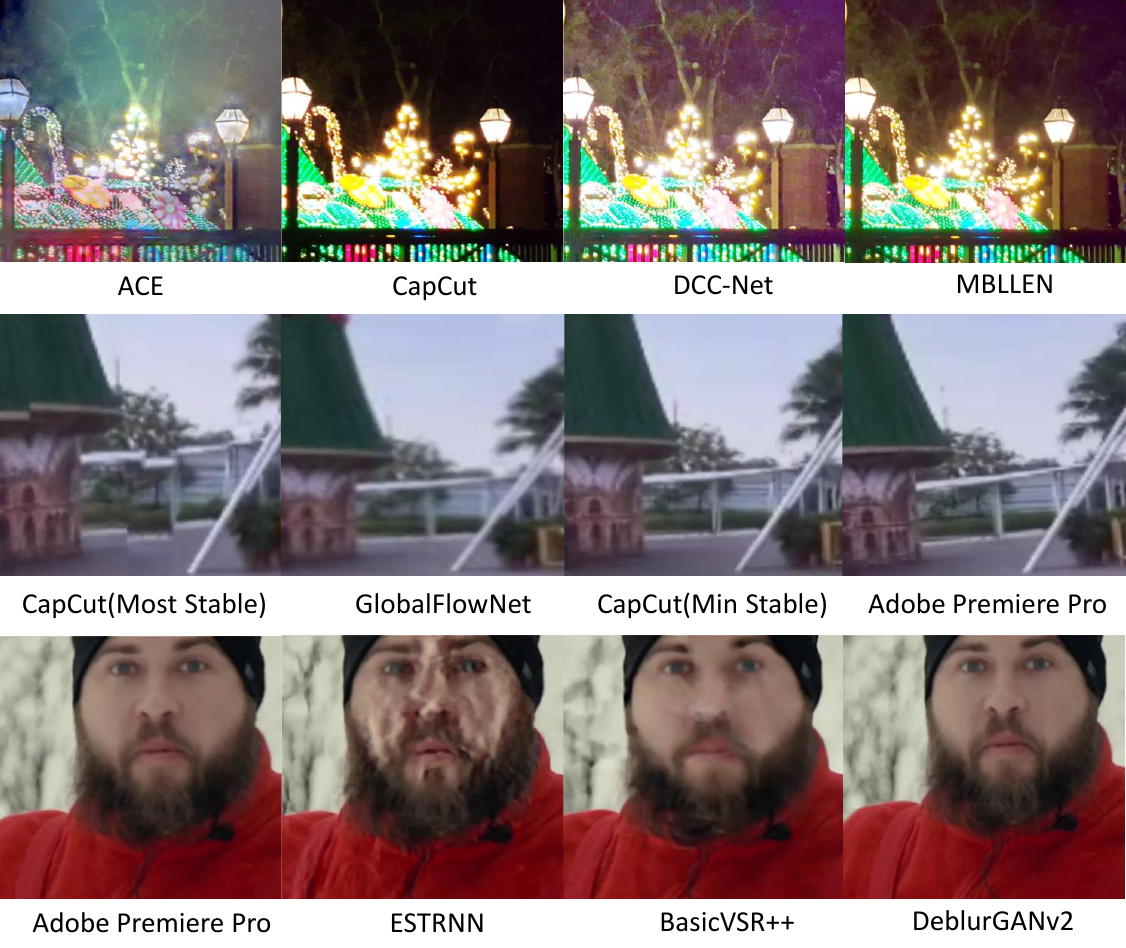}
    \caption{Example frames of the training sequences. The first row shows example frames corresponding to the videos generated using four different enhancement methods for color, contrast and brightness enhancement. The second row presents the examples from the content generated by four deshaking methods. The third row shows those produced by four deblurring methods.}
    \label{fig:Training Database Example}
\end{figure}

Each enhanced video and its corresponding reference counterpart are then randomly segmented with a non-overlapping spatio-temporal sliding window to produce training patches with the size of 256 (height) $\times$ 256 (width) $\times$ 3 (channel) $\times$ 72 (temporal length). The reference patches here are only used to generate pseudo-labels for quality classification and are not input into the model. This results in 13,156 (enhanced and reference) patch pairs. \cref{fig:Training Database Example} shows example frames of the training sequences generated for training the proposed model.

\subsection{Training Strategy}\label{contrastivelearning}

In self-supervised learning, when the model is difficult to optimize directly for its primary task, it is often trained to perform a related pretext task, which can be learned more effectively. In our case, since assessing the quality of enhanced video content is a challenging task and the corresponding labels are difficult to obtain, inspired by contrastive learning \cite{jaiswal2020survey,le2020contrastive}, we employ a projector network (with two layers of multilayer perceptron), $g(\cdot)$, to transform the original task into two classification problems focusing on content and quality classification, rather than predicting absolute quality indices.  

Specifically, the projector first takes the output of the RMViT, $\textbf{h}^v$, and obtains the quality representation of the video,  $\mathbf{z}\in \mathbb{R}^{128\times1}$,  which is expected to represent the quality of this sequence. This is used to calculate the quality-aware loss. On the other hand, we perform average pooling on the processed frame embeddings in the last recurrent iteration, $[\tilde{\mathbf{h}}_{T-N+1} \circ \tilde{\mathbf{h}}_{T-N+2} \circ ... \circ \tilde{\mathbf{h}}_{T}]$, to generate a content embedding, $\mathbf{h}^c \in \mathbb{R}^{2048\times1}$. We also feed the memory token generated in the second last recurrent iteration, $\mathbf{h}_\mathrm{mem}^{S-1}$,  into a prediction network \cite{CSPT}, $f(\cdot)$, to obtain the predicted content embedding $\hat{\mathbf{h}}^c$. Both $\mathbf{h}^c$ and $\hat{\mathbf{h}}^c$ are then passed to the same projector, $g(\cdot)$, to obtain their corresponding content representation, $\mathbf{c}\in \mathbb{R}^{128\times1}$, and the predicted content representation, $\hat{\mathbf{c}}\in \mathbb{R}^{128\times1}$, respectively, which are used to calculate the content-aware loss.

To enable contrastive learning, a batch ($2B$) of training patches are fed into the network with $B$ randomly selected patches of size 256$\times$256$\times$3$\times$72 and their corresponding down-sampled versions (128$\times$128$\times$3$\times$72). The latter is used here to provide true positive pairs (for the calculation of the quality-aware loss) in contrastive learning, as in \cite{contrique}. The contrastive loss function contains two components, quality-aware and content-aware losses.

\noindent\textbf{Quality-aware loss} focuses on distinguishing videos based on the similarity of their visual quality. For a given input patch, the positive pairs are constituted either between the patches with similar quality or between a full-resolution patch and its corresponding low-resolution counterpart. The quality-aware loss, $L_i^{quality}$, for the $i^{th}$ input patch in a batch is defined by:
\begin{equation}
\label{loss1}
\displaystyle
L_i^{quality} = - \frac{1}{P_i} \sum_{j =1}^{P_i} \log \left( \frac{\exp(\phi(\mathbf{z}_i, \mathbf{z}_j)/\tau)}{\sum_{k=1,k \neq i}^{2B} \exp(\phi(\mathbf{z}_i, \mathbf{z}_k)/\tau)} 
\right).
\end{equation}
Here $P_i$ represents the number of patches (positive pairs) in a batch with the same quality interval (the classification process is described below) as patch $i$, or with the similar content to patch $i$ but in different resolutions. \( \phi \) stands for the normalized dot product function, \( \phi(a, b) = a^Tb\ / (\|a\|_2 \|b\|_2)\), which measures the similarity between $a$ and $b$. $\tau$ is a temperature parameter that is less than 1.

\noindent\textbf{Quality classification.} To support quality-aware learning, we employ a proxy quality metric to perform quality classification, inspired by the ranking-based training methodology proposed in RankDVQA \cite{feng2024rankdvqa}.  Specifically, we first use VMAF \cite{w:VMAF} to calculate the quality score for each patch by comparing it with the associated corresponding reference (see \cref{ssec:DatabaseGeneration}). During the training process, for patch $i$, if another patch $j$ is associated with a VMAF value close to that of patch $i$ (the difference is smaller than a threshold $\mathrm{TH}$),
\begin{equation}
    | \mathrm{VMAF}_i-\mathrm{VMAF}_j|<=\mathrm{TH},
\end{equation}
these are considered as a positive pair. Similarly as in \cite{feng2024rankdvqa}, although VMAF \cite{w:VMAF} does not offer a perfect correlation performance with groundtruth subjective opinions, with a sensible threshold value, the classification here can be considered to be reliable (with a 95\%+ accuracy), as demonstrated in \cite{feng2024rankdvqa}). 

\noindent\textbf{Content-aware loss} captures the content-dependent nature of video enhancement methods, providing a different observation aspect in the training process. This has been reported to be effective in the literature \cite{CSPT} for the video quality assessment task. In our case, the positive pair from the content perspective is defined for patches with the same source content.

Specifically, given the content representation $\mathbf{c}_i$ and its predicted version $\hat{\mathbf{c}}_i$ for patch $i$ in a batch, the content-aware loss is calculated by:
\begin{equation}\label{loss2}
    L_i^{content} = -\frac{1}{C_i} \sum_{j=1}^{C_i}\log \left(\frac{\exp(\phi(\mathbf{c}_i, \mathbf{c}_j)/\tau)+\exp(\phi(\mathbf{c}_i, \hat{\mathbf{c}}_j)/\tau)}{\sum_{k=1,k \neq i}^{B}(\exp(\phi(\mathbf{c}_i, \mathbf{c}_k)/\tau) + \exp(\phi(\mathbf{c}_i, \hat{\mathbf{c}}_k)/\tau)) }\right),
\end{equation}
where $C_i$ represents the number of patches in a batch with the same content as patch $i$.

Similarly as in \cite{khosla2020supervised}, the final contrastive loss is defined as a weighted sum of the quality- and content-aware components:
 \begin{equation}\label{loss3}
 L = \frac{1}{B} \sum_{i=1}^{B} \left( L^{quality}_i + \lambda_1 L^{content}_i \right),
\end{equation}
where \( \lambda_1 \) is a tuning parameter. Here, we only consider full-resolution patches when calculating the final contrastive loss.

\section{Experimental Configuration}

\noindent\textbf{Implementation Details.} The implementation of the RMViT module is based on the Vision transformer for small datasets \cite{lee2021vision}, where the depth is set to 8 and the number of heads is 64. The hidden dimension is adjusted to 2048 following the practice in \cite{contrique}. The size of the segment is fixed at 4. The number of memory tokens and the segment length are 12 (this is verified in \textit{Ablation Study} through an analysis of the training results). The batch size for training is $B$ = 256. $\tau$ is set to 0.1. The threshold $\mathrm{TH}$, used in the quality classification, is set to 6 based on \cite{feng2024rankdvqa}. Data augmentation is performed during the training content generation through block rotation. The RMViT module, the prediction network, $f(\cdot)$, and the projector network, $g(\cdot)$, were trained simultaneously from scratch for 150 epochs using stochastic gradient descent (SGD) at a learning rate of 0.00025. A linear warm-up over the first ten epochs was applied to the learning rate, followed by a cosine decay schedule used in \cite{loshchilov2016sgdr}. The proposed method was implemented using Pytorch 1.13 with an NVIDIA 3090 GPU.

\noindent\textbf{Benchmark blind VQA methods.} The proposed quality metric is compared with ten blind VQA methods, including two conventional models, NIQE \cite{niqe}, BRISQUE \cite{brisque}, three regression-based methods, V-BLIINDS \cite{VBLIINDS}, TLVQA \cite{TLVQM} and ChipQA \cite{ChipQA}, and five deep learning based approaches, BVQA \cite{BVQA}, VSFA \cite{VSFA}, SimpleVQA  \cite{SimpleVQA}, CONVIQT \cite{conviqt}, FAST-VQA \cite{FastVQA} and RanKDVQA-NR \cite{feng2024rankdvqa}. It is noted that some well-performing no-reference VQA models, such as TB-VQA \cite{wu2023video}, which ranks first in the NTIRE 2023 quality assessment of video enhancement challenge \cite{liu2023ntire}, and SB-VQA \cite{huang2023sb}, have not been included in this experiment, as their source codes are not public accessible.

\noindent\textbf{Evaluation settings.} Due to the limited test content available, we performed a five-fold cross validation experiment based on the VDPVE database for RMT-BVQA and all the other eight learning-based (regression or deep learning) VQA methods. It should be noted that here we refer to the \textbf{VDPVE training set}, which contains 839 sequences, while the corresponding test set is not accessible, as mentioned in \cref{sec:LR}. 
This dataset is further divided into three sub-datasets: Subset A including 414 videos generated through colour, brightness, and contrast enhancement; Subset B comprising 210 deshaked videos; and Subset C containing 215 deblurred videos.  It should be noted that for the proposed method, only the linear regression is optimized using the training set in the cross validation. The five-fold cross validation has been performed 100 times (the split is based on source content) to calculate the average correlation coefficients. All trainable methods (with their pre-trained models) were fine-tuned for up to 150 epochs during each cross validation with an early stopping strategy. To test the correlation performance with subjective opinions, we employed both the SRCC and the Pearson Linear Correlation Coefficient (PLCC) as evaluation metrics. We have also compared the computational complexity of our approach with the other five deep learning-based VQA methods in terms of runtime and model parameters (see \cref{table: Model Complexity}).

\section{Results and Discussion}
\label{sec:results}

\begin{table}[t!]
\caption{Summary of the comparison and ablation study results. Here the best and second best results in each column are highlighted in \textcolor{red}{red} and \textcolor{blue}{blue} colours, respectively. \label{table:results}}
\begin{center}
\resizebox{0.95\textwidth}{!}{% Resize table to fit within the textwidth
\begin{tabular}{@{}r|c|c|c|c|c|c|c|c@{}}
\toprule
& \multicolumn{2}{c|}{Subset A} & \multicolumn{2}{c|}{Subset B} & \multicolumn{2}{c|}{Subset C} & \multicolumn{2}{c}{Overall} \\
\cmidrule(r){2-3} \cmidrule(l){4-5} \cmidrule(l){6-7} \cmidrule(l){8-9}
NR Metrics & SRCC & PLCC & SRCC & PLCC & SRCC & PLCC & SRCC & PLCC \\ 
\midrule
NIQE \cite{niqe}&  0.3555 &0.4485 &0.5830 &0.6108 &0.0540 &0.2079 &0.1401 &0.2411\\
VIIDEO \cite{VIIDEO}& 0.1468 &0.3484 &0.0854 &0.3387 &0.2701 &0.3104 &0.0646  &0.2574 \\\midrule
V-BLIINDS \cite{VBLIINDS}& 0.7214   & 0.7691 & 0.7028 & 0.7196 & 0.7055 & 0.7104 & 0.7106 & 0.7301\\
TLVQM  \cite{TLVQM}&  0.6942& 0.7085& 0.5619  &0.5940& 0.5457 &0.6001 &0.5861 &0.6499\\ 
ChipQA \cite{ChipQA}    & 0.4572 & 0.4756 & 0.3347 & 0.3753 & 0.7713 & 0.7759 & 0.5639 & 0.5285 \\
\midrule
BVQA \cite{BVQA}      & 0.5477 & 0.5596 &  0.3986 &  0.4271 & 0.3403 & 0.3872  & 0.4655 & 0.4807 \\
VSFA \cite{VSFA}      & 0.4803 & 0.4912 &  0.5315 & 0.5696 & 0.6564 &  0.6911 &  0.5282 & 0.5473 \\
CONVIQT \cite{conviqt}  & \textcolor{blue}{{0.7411}} & \textcolor{blue}{{0.7639}} & 0.4174 & 0.6926& 0.6678 & 0.7192 & 0.7052 & 0.7297 \\
FAST-VQA \cite{FastVQA}  & 0.7022& 0.7147 & \textcolor{blue}{{0.7398}} & \textcolor{blue}{{0.7706}} & \textcolor{red}{{0.8356}} & \textcolor{red}{{0.8677}} & \textcolor{blue}{{0.7196}} & \textcolor{blue}{{0.7644 }} \\
RankDVQA-NR \cite{feng2024rankdvqa} &0.6620 &0.6703 &0.6623 & 0.6527 & 0.5524 &0.4872 &0.6197 & 0.5777 \\

\midrule
{\textbf{RMT-BVQA (ours)}}  & \textcolor{red}{\textbf{0.8164}} & \textcolor{red}{\textbf{0.8139}} & \textcolor{red}{\textbf{0.8012}} & \textcolor{red}{\textbf{0.8019}} & \textcolor{blue}{\textbf{0.8315}} &\textcolor{blue}{\textbf{0.8385}} & \textcolor{red}{\textbf{0.8209}} & \textcolor{red}{\textbf{0.8384}} \\\midrule
v1-GRU & 0.7906 & 0.8012 &0.5585 &0.6085 & 0.8128 & 0.8209 &0.7852 &0.7880\\ 
v2-quality & 0.8019 & 0.8023 &0.7989& 0.8049 & 0.7864 & 0.8075 &0.8052 &0.8255\\
\bottomrule
\end{tabular}
}
\end{center}

\end{table}

\subsection{Overall performance}

\begin{figure}[!t]
    \begin{minipage}[b]{1\linewidth}
        \centering
        \tiny
        \centerline{\includegraphics[width=1\linewidth]{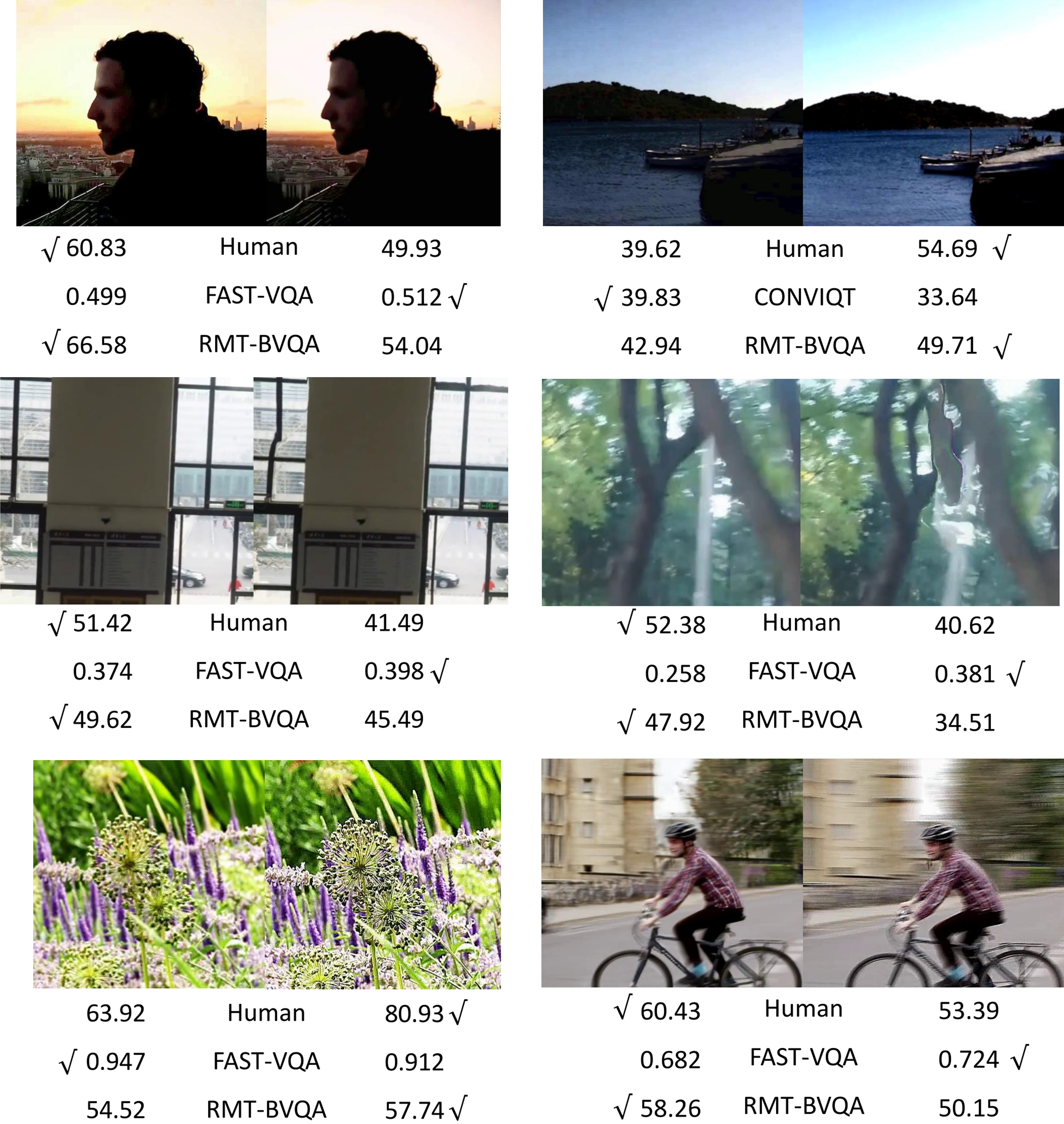}}
    \end{minipage}

    \caption {Visual examples from three subsets of VDPVE \cite{gao2023vdpve} (from the top to the bottom: color transform, deshaking and deblurring) demonstrating the superiority of the proposed method. In these three cases, RMT-BVQA (ours) provides the same quality differentiation as human perception does.}
    \label{fig:visual}
\end{figure}

\cref{table:results} summarizes the comparison results between our proposed method, RMT-BVQA, and ten benchmark VQA methods. It should be noted that the results presented here are different from \cite{liu2023ntire}, where the test set (unavailable publicly, as mentioned in in \cref{sec:LR}) of the VDPVE database was employed for evaluation. In this experiment, we performed cross validations (100 times) on the VDPVE training set instead. It can be observed in \cref{table:results} that RMT-BVQA offers the best overall performance among all the tested quality metrics - the only one with SRCC and PLCC values above 0.8. For all three subsets, A, B, and C in the VDPVE database, which correspond to different enhancement method types, our model is the best performer on Subset A and B, and the second best on Subset C (very close to FAST-VQA). \cref{fig:visual} provides three visual examples in which the proposed RMT-BVQA offers the correct quality differentiation as human perception, while the second (FAST-VQA \cite{FastVQA}) or the third best performer (CONVIQT \cite{conviqt}) fails to do that.

\subsection{Ablation study}

To further verify the effectiveness of the main contributions in this work, an ablation study is also conducted including the following sub-tests.

\noindent\textbf{Recurrent Memory Vision Transformer.} We evaluated the contribution of our RMViT module by replacing it with an alternative network structure, Gated Recurrent Unit (GRU) \cite{cho2014properties}, which has been utilized in previous contrastive learning based video quality assessment tasks \cite{conviqt}. The new variant is denoted by (v1-GRU), and its results (based on the same cross-validation experiment) are shown in \cref{table:results}. It can be observed that v1-GRU achieves lower average correlation coefficients compared to the full RMT-BVQA, in particular on Subset B (deshaking). This verifies the contribution of the recurrent memory vision transformer, due to its long-term scene memory compared to the traditional GRU module, especially for those videos containing severe scene transitions (in Subset B). 

Moreover, to determine the values of two hyperparameters in this RMViT module, including the number of memory tokens, $M$, and the segment length, $S$ (used in training), we performed an analysis based on the training performance by varying these values within predefined ranges (limited by our memory constraints), $2\leq M\leq12$ and $2\leq S\leq12$. As shown in \cref{tab:RMViT Analysis}, when both the length of segments and the number of memory tokens are set to 12 (we cannot test higher values due to memory constraints), the training loss reaches its lowest level. This justifies the selection of these two hyperparameter values.

\begin{table}[!t]
    \caption{The analysis on the memory token size, $M$ and the segment length $S$, used in the training process. Here the size of the red circle indicates the value of the lowest training loss.}
    \label{tab:RMViT Analysis}
    \centering
        \begin{tabular}{p{0.5cm}| >{\centering\arraybackslash}p{0.7cm} |>{\centering\arraybackslash}p{1.2cm} >{\centering\arraybackslash}p{1.2cm}  >{\centering\arraybackslash}p{1.2cm} >{\centering\arraybackslash}p{1.2cm} >{\centering\arraybackslash}p{1.2cm} >{\centering\arraybackslash}p{1.2cm}  }
    \toprule
     \multicolumn{2}{c|}{} & \multicolumn{6}{c}{The number of memory tokens $M$}\\
    \cmidrule{3-8}
         \multicolumn{2}{r|}{Loss}  & 2 & 4 & 6 & 8 & 10 & 12 \\
    \midrule    
    \multirow{6}{*}{\rotatebox[origin=c]{100}{$S$}} 
        &12& 6.236 & 5.927 & 5.728 & 5.475 & 5.151 & \textcolor{red}{\textbf{5.133}}\\
        &10 & 6.374 & 5.762 & 5.783 & 5.564 & 5.174 & 5.166\\
         & 8&  6.490  &  6.271 &  5.949 &  5.770 & 5.569 & 5.492\\
         &6& 6.698 & 6.487 & 6.174 & 6.052 & 5.961 & 5.944 \\
         &4& 6.765 & 6.519 & 6.434 & 6.326 & 6.309 & 6.341 \\
         &2& 6.889 & 6.879 & 6.747 & 6.654 & 6.541 & 6.553 \\
         \bottomrule
    \end{tabular}%
    
\end{table}

\noindent\textbf{Content-quality-aware contrastive learning.} We further verify the effectiveness of the proposed content-quality-aware loss function by removing the content-aware loss, producing (v2-quality). We did not test the contribution of the quality-aware loss, because only employing content-aware loss leads to unstable training performance. The results in \cref{table:results} show that v2-quality is worse than the original RMT-BVQA, which confirms the effectiveness of the content-quality-aware contrastive learning methodology.

\noindent\textbf{The training database.} Since it is difficult to find an alternative to the proposed training database, which can support the optimization of the RMViT module, we instead compared v1-GRU to the original CONVIQT model to verify the contribution of the training database. v1-GRU effectively has the same network architecture as CONVIQT - differing only in the use of the new training database to optimize the GRU module. From \cref{table:results}, we can observe that v1-GRU outperforms CONVIQT, with better overall correlation performance - this justifies the contribution of the developed training database. 

\subsection{Model Complexity}
\cref{table: Model Complexity} provides the computational complexity results for the proposed method and the other five deep learning based benchmark methods, in terms of runtime (second, for processing 300 frames 1280$\times$720 video) and the number of parameters. This experiment is implemented on a workstation with an NVIDIA 3090 GPU, a 3.30 GHz Intel W-1250 CPU and 64 GB of RAM. The high complexity in particular in model size is mainly due to the complex network structure employed (vision transformer in the RMViT module). Although our parameters are significantly higher compared to the benchmarks, the runtime has not increased proportionally, showing only a 30.48\% increase compared to CONVIQT \cite{conviqt}. This discrepancy arises from considering only the memory tokens generated by the final two segments during inference, thereby substantially reducing the model's effective runtime.
%-------------------------------------------------------------------------

\begin{table}[!t]
    \caption{The runtime and model size (number of parameters) figures for the proposed method and the other learning-based benchmarks.}
    \label{table: Model Complexity}
    \begin{center}
    \small
    \begin{tabular}{r|r|r}
    \toprule
        Benchmark & \ \ \ \ Runtime (s)& \ \ \ \ \  Params (M) \\ \toprule
        BVQA \cite{BVQA}   & 13.9 & 57.1\\\midrule 
        VSFA \cite{VSFA}    & 6.4  & 0.5 \\\midrule 
        CONVIQT \cite{conviqt}  & 18.7 & 38.6 \\\midrule 
        \ \ \ \ \ \ \ FAST-VQA \cite{FastVQA}  & 1.4& 27.6 \\\midrule
        RankDVQA-NR \cite{feng2024rankdvqa}  & 46.2& 4.6 \\\midrule 
        \textbf{RMT-BVQA} & 24.4 & 703.7 \\
    \bottomrule  
    
    \end{tabular}
    \end{center}
\end{table}

\section{Limitations of the proposed method}

Although our proposed method shows superior performance over the other benchmarked quality metrics, it is also associated with higher complexity, in particular in model size (\cref{table: Model Complexity}). This could potentially lead to greater energy consumption and a larger carbon footprint when implemented in practical applications. This issue can be alleviated by integrating dynamic input token pruning \cite{raposo2024mixture, wang2022efficient, kim2022learned}, into the RMViT module. Approaches such as knowledge distillation \cite{hinton2015distilling} and model compression \cite{buciluǎ2006model} can also be applied to further reduce the complexity of the model while maintaining performance. Moreover, the temporal pooling method employed in RMT-BVQA can be improved through more advanced feature fusion models \cite{gotz2021konvid}.

\section{Conclusion}
\label{sec:conclusion}
 In this paper, we propose a deep blind VQA method specifically for enhanced video content based on a novel self-supervised learning training methodology. In this work, we designed a new Recurrent Memory Vision Transformer (RMViT) module to obtain video quality representations, which is optimized through contrastive learning based on a content-quality-aware loss function. A large and diverse training dataset has also been developed containing various types of enhanced video content, which supports the proposed contrastive learning methodology but does not rely on expensive subjective tests to obtain ground-truth quality labels. The proposed method, RMT-BVQA, has been tested on a video enhancement quality database through five-fold cross validation, and exhibits higher correlation with opinion scores when compared to ten existing no-reference VQA methods. Future work should focus on benchmarking (and training) on more diverse enhanced video content and reducing the high complexity in particular in model size (\cref{table: Model Complexity}). The prediction performance of the proposed method on deblurred content (Subset C) can also be further improved.

% BibTeX users should specify bibliography style 'splncs04'.
% References will then be sorted and formatted in the correct style.
%

\section*{Acknowledgements}
Research reported in this paper was supported by an Amazon Research Award, Fall 2022 CFP. Any opinions, findings, and conclusions or recommendations expressed in this material are those of the author(s) and do not reflect the views of Amazon. The authors also appreciate the funding from the UKRI MyWorld Strength in Places Programme (SIPF00006/1).

\bibliographystyle{splncs04}
\bibliography{main}
\end{document}